\documentclass{pasj00}

\begin{document}
\SetRunningHead{Y. Takeda et al.}{Solar rotation inferred from radial 
velocities of the sun-as-a-star during eclipse}
\Received{2014/07/04}
\Accepted{2014/10/28}

\title{Solar rotation inferred from radial velocities of \\
the sun-as-a-star during the 2012 May 21 eclipse
\thanks{Based on data collected at Okayama Astrophysical Observatory (NAOJ, Japan).}
}

%

\author{
Yoichi \textsc{Takeda,}\altaffilmark{1,2}
Osamu \textsc{Ohshima,}\altaffilmark{3}
Eiji \textsc{Kambe,}\altaffilmark{4}
Hiroyuki \textsc{Toda,}\altaffilmark{4}
Hisashi \textsc{Koyano,}\altaffilmark{4}\\
Bun'ei \textsc{Sato,}\altaffilmark{5}
Yasuhisa \textsc{Nakamura,}\altaffilmark{6}
Norio \textsc{Narita,}\altaffilmark{1,2}
 and
Takashi \textsc{Sekii}\altaffilmark{1,2}
}

\altaffiltext{1}{National Astronomical Observatory, 2-21-1 Osawa, 
Mitaka, Tokyo 181-8588}
\email{takeda.yoichi@nao.ac.jp}
\altaffiltext{2}{The Graduate University for Advanced Studies, 
2-21-1 Osawa, Mitaka, Tokyo 181-8588}
\altaffiltext{3}{Mizushima Technical High School, 1230 Nishi-Achi,
Kurashiki, Okayama, 710-0807}
\altaffiltext{4}{Okayama Astrophysical Observatory, National 
Astronomical Observatory of Japan, \\
Kamogata, Okayama 719-0232}
\altaffiltext{5}{Tokyo Institute of Technology, 2-12-1 Ookayama, 
Meguro-ku, Tokyo 152-8550}
\altaffiltext{6}{Fukushima University, 1 Kanayagawa, 
Fukushima, Fukushima 960-1296}
%

\KeyWords{
line: profiles --- stars: eclipses --- Sun: rotation --- 
techniques: radial velocities --- techniques: spectroscopic} 

\maketitle

\begin{abstract}
With an aim to examine how much information of solar rotation can be 
obtained purely spectroscopically by observing the sun-as-a-star
during the 2012 May 21 eclipse at Okayama Astrophysical Observatory,
we studied the variation of radial velocities ($V_{\rm r}$), which were 
derived by using the iodine-cell technique based on a set of 184 high-dispersion 
spectra consecutively obtained over the time span of $\sim 4$~hours.
The resulting $V_{\rm r}(t)$ was confirmed to show the characteristic 
variation (Rossiter--McLaughlin effect) caused by time-varying visibility
of the solar disk.
By comparing the observed $V_{\rm r}(t)$ curve with the theoretical ones,
which were simulated with the latitude ($\psi$) dependent solar rotation 
law $\omega_{\rm sidereal}(\psi)$ = $A + B \sin^{2}\psi$  (deg day$^{-1}$), 
we found that the relation $B \simeq -5.5 A + 77$ gives the best fit, 
though separate determinations of $A$ and $B$ were not possible. 
Since this relationship is consistent with the real values known 
for the sun ($A \simeq 14.5$, $B \simeq -2.8$), we may state that our 
analysis yielded satisfactory results. This consequence may provide 
a prospect of getting useful information on stellar rotation of eclipsing binaries
from radial-velocity studies during eclipse, if many spectra of sufficiently 
high time-resolution are available. 
\end{abstract}

%


\section{Introduction}

It is almost 90 years since the peculiar behavior in the radial 
velocity curve of eclipsing binaries during eclipse was discovered
by the pioneering studies of Rossiter (1924) (for $\beta$~Lyr) and 
McLaughlin (1924) (for $\beta$ Per). They called this phenomenon as 
the ``rotational effect,'' since it is caused by the combination of 
two factors that (1) the visible part of the stellar disk 
changes during the course of eclipse, and (2) each point of the  
disk has its own line-of-sight velocity due to stellar rotation.

More specifically, in a typical eclipsing binary system where the rotation 
and orbital motion are in the same direction, this effect may be 
briefly described as follows:\footnote{
In this description, we assume that the radial velocity of the whole
eclipsing binary system ($\gamma$ velocity) is zero.}\\
--- (i) At the ingress phase, the radial velocity (starting from 
net-zero when all disk is visible) is generally positive and systematically 
increases until the maximum value is attained just before the mid-eclipse, 
since the relative contribution of the receding limb (where the 
line-of-sight velocity is maximum) becomes more important as the 
obscured fraction of remaining part of the disk progressively increases. \\
--- (ii) At the egress phase of eclipse, on the contrary, starting from 
the most negative value (caused by the approaching limb just re-appeared 
after mid-eclipse), the net (negative) radial velocity again gradually 
increases with time as the visible fraction of remaining part 
of the disk progressively increases, until it eventually ends with 
net-zero when the whole disk is visible.\\
--- (iii) At the mid-eclipse near the light minimum between (i) and (ii),
a conspicuous ``leap'' is observed, since the radial velocity abruptly 
changes its sign here from positive to negative.  

This effect, which is occasionally called as ``Rossiter--McLaughlin effect''
(hereinafter referred to as ``R--M effect'') after the discoverers, recently 
received hot attention of astronomers especially in connection with 
observations of transiting planets, since it can be applied to diagnose 
the properties (e.g., rotational velocity, axis of rotation, etc.) of 
not only the host star but also the planetary system (see, e.g., Ohta et al. 2005).

Yet, when it comes to eclipsing binaries belonging to the field of stellar 
physics, studying the properties of the eclipsed component by actively utilizing 
the R--M effect seems to have been rarely tried so far, despite its 
conceptual potentiality (e.g., Kallrath \& Milone 2009). 
Is this effect practically applicable to investigating the rotational 
features (e.g., rotational velocity, degree of differential rotation, etc.) 
of eclipsing binaries by analyzing the observed radial-velocity curve 
during the phase of eclipse?

We recently came to an idea of examining this possibility by making use 
of the unusual occultation phenomenon timely occurred in the exclusively 
well studied and most familiar star, our sun, since a deep (even if partial) 
solar eclipse by the moon took place in the morning of 2012 May 21 at Okayama 
Astrophysical Observatory, where sufficient experiences of precise radial 
velocity determinations by using the 188~cm reflector+ HIDES spectrograph 
have been accumulated over more than 10 years thanks to the long-continuing 
``Okayama Planet Search Program'' targeting evolved G--K giants 
(e.g., Sato et al. 2008). This situation motivated us to challenge 
a new spectroscopic study on this occasion.

Thus, following our research plan of this line, we first obtained the 
observed radial velocity curve of ``the sun as a star'' during the course 
of the eclipse by using HIDES, and then analyzed it to get information 
on the solar rotational properties as much as possible. The comparison of 
such obtained results with the actual solar parameters (which we already 
know very well) would make an interesting and valuable touchstone for 
discussing the possibility and the prospect of the R--M effect mentioned 
above. The purpose of this article is to report the consequence of 
this experiment.

We note that Molaro et al. (2013) recently informed a successful detection
of the R--M effect based on their high-precision radial-velocity observations
of integrated sunlight reflected by the moon during the Venus passage 
on 2012 June 6. While this study is similar to theirs (though the transiting
object is not Venus but the moon in our case), our aim is different
in the sense that it is not only to detect the R--M effect but also 
to extract therefrom as much information of solar rotation as possible.


\section{Observational data}

\subsection{Observation and spectrum reduction}

The detailed data of the partial solar eclipse observed at Okayama 
Astrophysical Observatory (OAO: longitude = 
E~133$^{\circ}$35$'$38$''$.24, latitude = 
+34$^{\circ}$34$'$34$''$.12,  altitude = 370~m) in the morning of 
2012 May 21 (JD~2456068.39--2456068.49
or from $\sim$~6:20~JST to $\sim$~8:50~JST 
with the duration of $\sim 2.5$~hours) are graphically described 
in figure~1 (positional changes of the sun and moon, variation in
the magnitude of eclipse with time, etc.), and the appearances of the sun
at representative phases are schematically depicted in figure 2.
All these positional data were generated with the help of HORIZONS\footnote{
$\langle$http://ssd.jpl.nasa.gov/horizons.cgi$\rangle$.}
operated by NASA/JPL. 
The aspect angles of the solar rotation axis ($P$ : position angle 
between the geographic north pole and the solar rotational north pole; 
$B_{0}$: heliographic latitude of the central point of the solar disk) 
on this day were $P = -19^{\circ}.1$ and $B_{0} = -1^{\circ}.95$.
In this morning, several appreciable active regions (spots, plages, 
dark filaments, prominences) were observed on the solar surface, 
as we can see in the H$\alpha$ filtergram (corresponding to the 
H$\alpha$ center) inserted in figure 1, which were taken on 
2012 May 21 (at 00:00:36 UT) just after the ending of eclipse by SMART 
(Solar Magnetic Activity Research Telescope) at the Hida Observatory 
of Kyoto University.\footnote{
The original image file (filename:  
``halpha\_p00\_20120521000036.jpg'') is available at
$\langle$http://www.hida.kyoto-u.ac.jp/SMART/pub/2012/05/21/T1/jpeg/$\rangle$.
}

Our spectroscopic observations of the eclipsed sun were carried
out without using the telescope. That is, the bare end of the
optical fiber was directly exposed to the solar radiation, so that
the collected light was further sent to HIDES (HIgh Dispersion
Echelle Spectrograph) via the fiber link system (Kambe et al. 2013).
The FOV (field of view) of our observation is a circle area with 
angular diameter of $\sim 30$~deg (corresponding to 0.21 str;
see the explanation below) centered on the sun. The practical setup of 
our equipment is described in the following, while graphically depicted 
in figure 3a (photographic view) and figure 3b (schematic optical layout).
\begin{itemize}
\item
Before the bare input end of the optical fiber 
(FBP100120140, Polymicro Co. Ltd.;
NA (numerical aperture) = 0.22 or f/2.2)
toward the sun, we put an iodine cell\footnote{
This is a copy of the one described in Kambe et al. (2002).} 
with its vacuum vessel for precise radial velocity measurement.
\item 
As a diffusor, a polishing paper with grain size of 1~$\mu$m was
inserted between the iodine cell and the fiber. This reduced
the fiber modal noise considerably as well as decreasing
the sunlights by less than one thousandth. 
\item
The FOV of the diffusor was restricted to about $\sim 0.21$~str
by the vacuum vessel's window, which allowed to keep monitoring
the whole sun disk under somewhat poor guiding 
while considerably limiting the scattered light coming directly
from the sky.
\item 
A hole of the filter box (or more exactly, in an alminum plate 
just after the vaccum vessel of the iodine cell) with diameter
of 17~mm worked as an entrance aperture to the fiber.
This assures that the F-ratio of the incident light (f/3.8)
is close to that of our normal observations with the telescope
and the fiber system.
\item 
These fiber-entrace components were put on a portable equatorial 
telescope mount (Vixen GP.D) for automatically tracking the sun.
\end{itemize}

Practically, as we did not have sufficient time to set up the telescope 
mount properly between the end of the regular night observing run and 
the solar eclipse, we had to compensate the tracking error manually 
from time to time.
For this purpose, we put pinhole and target plates on the same 
mount, which were separated by about 200~mm with their axes being 
aligned to the fiber input end beforehand.
We guided the sun so that the solar image on the target plate did not 
move by more than 15 arcmin.
At the output end of the optical fiber (i.e., at the entrace of 
the spectrograph), a microlens was used to convert the F-ratio of 
the incident light to that of the spectrograph (F/29),  and then 
an image slicer was placed to keep a moderate wavelength resolving
power. This optics is identical to that of our HIDES fiber-link 
(Kambe et al. 2013). In this way, we could obtain spectra almost 
equivalent to those of the sun seen as a star (solar flux spectra).

Although the observations were done over the time span of $\sim 4$~hours
(JD~2456068.35--2456068.51 or 5:20--9:20 JST) 
covering the eclipse, the sky condition was not always desirable,
especially before/around the beginning of eclipse where the
observations were often disturbed due to clouds. 
Just after the eclipse began (from $\sim$~JD 2456068.4), however, 
the condition was significantly improved; since then, we could conduct
our observations almost continuously and satisfactorily 
(even if thin cirrus passed by sometimes). 
Consequently, we could obtain 184 spectra (with I$_{2}$ lines
imposed mainly at $\sim$~5000--6000~$\rm\AA$) covering 4400--7500~$\rm\AA$ 
with three mosaicked CCDs) with a wavelength resolving power of $R = 52000$. 
The exposure time for each spectrum frame was set to $\sim 10$~s in most cases 
(while the overhead time for CCD readout was $\sim 40$~s).

The reduction of the spectra (bias subtraction, scattered-light correction,
aperture determination, flat-field correction, spectrum extraction,
wavelength calibration, continuum normalization, etc.) was carried out
by using the {\tt echelle} package of IRAF.\footnote{IRAF is distributed 
by the National Optical Astronomy Observatories, which is operated by the 
Association of Universities for Research in Astronomy, Inc. under cooperative 
agreement with the National Science Foundation, USA.}. 
The typical signal-to-noise ratios (S/N) of the finally resulting spectra are 
$\sim$ 400--500 
(between minimum $\sim 200$ and maximum $\sim 800$),  
as estimated from photon counts at $\lambda \simeq 5500$~$\rm\AA$. 
As an example, we show in figure 4 one of our typical spectra (taken at just
outside of the eclipse) in two wavelength regions of 5335--5375~$\rm\AA$ 
(conspicuous I$_{2}$ lines) and 6685--6725~$\rm\AA$ (absence of I$_{2}$ lines)

As we can see from figure 4, our HIDES spectrum appears similar to 
Kurucz et al.'s (1984) solar flux spectrum, except that lines in our spectrum 
are slightly shallower because of the difference in the resolving power.
As a more quantitative check, we measured the equivalent widths [$EW$(OAO)] of
$\sim 100$ (almost blend-free) Fe~{\sc i} and Fe~{\sc ii} lines on our spectrum 
in $\sim$~6200--7400~$\rm\AA$, and compared with those $EW$(KPNO) measured from
Kurucz's (1984) KPNO spectrum atlas, which were already published in Takeda et al. 
(2005; cf. table~E1 therein). We found a fairly good correlation between
$EW$(OAO) and $EW$(KPNO), as represented by a linear regression
relation of $EW$(OAO) = 0.963 $EW$(KPNO) + 1.59 (where $EW$ is measured
in m$\rm\AA$) with a correlation coefficient of $r = 0.99$. Accordingly, 
our spectrum may be regarded practically as that of the sunlight integrated
over the disk.

\subsection{Radial-velocity determination}

The radial velocity of the sun-as-a-star for each frame was derived 
from the spectrum of $\sim$~5000--6000~$\rm\AA$ region (where many 
reference lines of I$_{2}$ molecules are manifestly imprinted) 
basically following the standard procedure (e.g., Sato et al. 2002), 
while using the sky spectrum taken on 2012 January 17 as the template 
spectrum.\footnote{
In radial-velocity determinations based on the I$_{2}$-cell technique, 
only the differential radial velocity of the spectrum is obtained 
relative to an arbitrarily adopted template spectrum. In this sense, 
only their time variations are of significance, while the absolute 
values themselves do not have much meaning.
} 
The spectrum of the relevant wavelength range was divided into segments 
of several~$\rm\AA$, and $\tilde{v}_{i}$ (radial velocity
relative to the template) was determined for each segment $i$ 
($i$ = 1, $\cdots$, $N$).
Then, the radial velocity of the sun-as-a-star relative to that of 
the template ($V_{\rm rel}$) and its probable error ($\epsilon$) were 
computed by averaging these $\tilde{v}_{i}$ over the segments as follows:
\begin{equation}
V_{\rm rel} \equiv \frac{\sum_{i=1}^{N} \tilde{v}_{i}}{N}
\end{equation}
and
\begin{equation}
\epsilon \equiv \sqrt{\frac{\sum_{i=1}^{N}(\tilde{v}_{i} - V_{\rm rel})^{2}}{N(N-1)}}.
\end{equation}

The actually used segments for computing the final result and its error 
with equations (1) and (2) were carefully chosen by excluding those showing 
large deviation from the main trend; and the number of 
finally adopted segments was $N = 196$ (about half of the total).

Here, given that only relative time variations are relevant in the present study
(cf. footnote 6), we can add any constant to $V_{\rm rel}$ (relative velocity 
in reference to the template). Actually, these $V_{\rm rel}$ values (i.e.,
velocity differenes between Jan. 17 and May 21) turned out to be rather large, 
which may be ascribed either to the large time span between observed dates or 
to some systematic effect because the template spectrum was obtained in 
the normal mode of HIDES using the 188~cm telescope (i.e., different optical 
system from that used in this study). 
Since the time average of $V_{\rm rel}$ over the whole period is 
$\langle V_{\rm rel}\rangle = 845.5$~km~s$^{-1}$, we apply an offset by this amount 
and hereafter use $V_{\rm r}^{\rm obs} (\equiv V_{\rm rel} - 845.5$~km~s$^{-1}$) 
as the formal observed radial velocity, which is convenient because 
$V_{\rm r}^{\rm obs}$ outside the eclipse takes values nearly zero.

The resulting radial velocities $V_{\rm r}^{\rm obs}$ are presented 
in electronic table~E (tableE.dat) for all spectra, and how they varied 
during the course of eclipse is displayed in figure 5, where the related 
quantities such as S/N and $\epsilon$ are also shown. 
It is evident from figure 5d that the variation of $V_{\rm r}^{\rm obs}$
has a characteristic shape of the R--M effect (though the sense of change is 
contrary to what was explained in section 1, since the occultation began 
from the side of the receding limb in this case).  

Regarding the error involved in $V_{\rm r}^{\rm obs}$, we can see from 
figure 5c that $\epsilon$ is between $\sim 5$~m~s$^{-1}$ and 
$\sim 10$~m~s$^{-1}$, which is about twice as large as that we have 
normally accomplished for similar S/N in our planet search observations.
We also note a correlation between $\epsilon$ and S/N in the sense 
that $\epsilon$ tends to increase at S/N~$\ltsim 300$ (cf. figure 6).
Since such cases of comparatively lower S/N correspond to
the phases of dimmed sunlight (before the eclipse disturbed by clouds,
or near to the mid-eclipse; cf. figure 5), we speculate that 
differences between the actually observed integrated spectrum and the 
adopted template sky spectrum may be responsible for this tendency of
somewhat larger $\epsilon$ (i.e., intrinsic changes of line profiles 
must have occurred because light contributions from various parts of the disk 
were inhomogeously integrated in such cases). At any event, $\epsilon$ 
does not have any significant influence\footnote{
The appreciable fluctuation observed in figure 5d before the 
beginning of the eclipse (at $\sim$~JD 2456068.37--2456068.38) is not 
due to errors in $V_{\rm r}^{\rm obs}$. This is most probably due to 
the fact that lights from various parts of the solar disk were 
inhomogeneously/irregularly contributed in the integrated observed flux, 
since clumpy clouds were incessantly passing through in front of 
the sun at this period. 
}
on the essential behavior of $V_{\rm r}^{\rm obs}$,
as we can see in figure 5d where the error bars are hardly recognized.
This observed $V_{\rm r}^{\rm obs}$ curve will be compared 
with the simulated theoretical $V_{\rm r}^{\rm cal}$ in section 4, 
the computational details for which are described in the next section 3.

\section{Modeling of radial velocity}

We assume that the sidereal angular rotational velocity of the sun 
($\omega_{\rm sidereal}$) depends on the heliographic latitude ($\psi$) as
\begin{equation}
\omega_{\rm sidereal} = A + B \sin^{2} \psi
\end{equation}
where $A$ and $B$ (in deg~day$^{-1}$) are the parameters corresponding to 
the equatorial rotation velocity and the degree of differential rotation, 
respectively.\footnote{
This two-parameter form is usually adopted to express solar differential rotation 
determined by the sunspot-tracing method. In case of using the Doppler method
where rotation can be investigated up to high-latitude zones, the
three-parameter form (with an additional term of $C \sin^{4} \psi$)
is often applied (see, e.g., Takeda \& Ueno 2011). Here, we decided to
adopt the former simpler one.}

Since $x$--$y$ Cartesian coordinate system is defined on the solar disk
in such a way that $y$ axis is aligned with the rotational axis 
(cf. figure 7), the following relation between $\psi$, $x$, and $y$ holds 
($R$ is the solar radius):
\begin{equation}
R \sin \psi = \sqrt{R^{2} - x^{2} - y^{2}}\sin B_{0} + y \cos B_{0}.
\end{equation}

Then, as the line-of-sight velocity due to solar rotation $v_{\rm los}$ 
at $(x,y)$ is expressed as
\begin{equation}
v_{\rm los} = x \cos B_{0} \omega_{\rm sidereal},
\end{equation}
we can calculate $v_{\rm los}$ at any point on the solar disk
with the help of equations (3), (4), and (5), if $(A, B)$ are specified.

This $v_{\rm los} (x, y)$ (in the heliocentric system) is then 
corrected by subtracting the heliocentric correction $\Delta^{\rm hel}(x,y,t)$ 
due to earth's motion (not only disk-position-dependent but also time-dependent; 
computed by following the procedure described in subsection 4.1 of 
Takeda \& Ueno 2011)\footnote{
Namely, $\Delta^{\rm hel}(x,y,t)$ was evaluated 
by using the ``rvcorrect'' task of IRAF, while taking into account (i) the 
rotation of the earth (diurnal velocity), (ii) the motion of the earth's center 
about the earth--moon barycenter (lunar velocity), and (iii) the motion of 
the earth--moon barycenter about the center of the Sun (annual velocity).
Practically, the heliocentric correction $\Delta^{\rm hel}(x,y,t)$ (for 
the observation at time $t$ on the disk point $(x, y)$) was computed 
at the coordinate ($X_{0}(t) + x, Y_{0}(t) + y$) on the celestial sphere,
where $X_{0}(t)$ and $Y_{0}(t)$ (the disk-center position of the Sun) 
were obtained by interpolating the solar ephemeris table. 
Note that this $\Delta^{\rm hel}(x,y,t)$ can not be simply represented 
by the disk-center value of $\Delta^{\rm hel}(0,0,t)$, since it 
significaltly depends on the position of the solar disk.
}
to obtain the topocentric value 
$v_{\rm los}^{\rm topo} (x, y, t)$ corresponding to our observation at Okayama:
\begin{equation}
v_{\rm los}^{\rm topo} (x, y, t) \equiv v_{\rm los} (x, y) - \Delta^{\rm hel}(x,y,t).
\end{equation}
Consequently, $V_{\rm r}^{\rm cal}$ (simulated radial velocity of the 
eclipsed sun) is evaluated by averaging $v_{\rm los}^{\rm topo} (x, y, t)$
over $D_{\rm vis}(t)$ (visible part of the solar disk) while weighting it 
according to the brightness at each disk point:
\begin{equation}
V_{\rm r}^{\rm cal} (t) = \frac{\iint_{D_{\rm vis}(t)} v_{\rm los}^{\rm topo} (x, y, t) I(x,y) dxdy}
       {\iint_{D_{\rm vis}(t)} I(x,y) dxdy},
\end{equation}
where the numerical integration was performed by dividing the solar disk into 36000 
($\equiv 360 \times 100$: with steps of 1$^{\circ}$ and 0.01$R$) tiny sections (figure 7). 
Regarding the disk brightness $I(x,y)$, we adopted a linear limb-darkening relation:
\begin{equation}
I(\mu) = I_{0} (1 - \epsilon + \epsilon \mu),
\end{equation}
where $I_{0}$ is the specific intensity at the disk center, 
$\mu$ is the direction cosine ($\mu \equiv 1 - \sqrt{x^{2}+y^{2}}/R$),
and $\epsilon$ is the limb-darkening coefficient assumed to be 0.7
(corresponding to $B-V \sim 0.65$ and $\lambda \sim$~5000--6000~$\rm\AA$;
cf. Fig. 17.6 in Gray 2005).\footnote{We examined the effect of changing $\epsilon$
by $\pm 0.1$ on the radial velocity curve, and found that the resulting variation 
was practically negligible (i.e., $\sim$~6--7~m~s$^{-1}$ at most).} 

In this way, we could compute $V_{\rm r}^{\rm cal}$ at any phase of eclipse for 
various combinations of $(A, B)$. Several examples of $V_{\rm r}^{\rm cal}(t)$ curves are 
depicted in figure 8a (changing $A$ while $B$ is fixed at 0) and figure 8b (changing 
$B$ while $A$ is fixed at 14.0).

\section{Result and discussion}

\subsection{Solution search for solar $A$ and $B$}

We are now ready to extract information of ($A$, $B$) 
by comparing $V_{\rm r}^{\rm cal}$ with $V_{\rm r}^{\rm obs}$.
For this purpose, we first computed 441 radial velocity curves
 $V_{\rm r}^{\rm cal}(t_{i})$ ($i=1, 2, \ldots, N$; $N = 184$ 
is the number of available spectra for each of the observing times),
resulting from combinations of 21 $A$ values (from 12.0 to 16.0 with a step of 0.2)
and 21 $B$ values (from $-5.0$ to +5.0 with a step of 0.5).
Then, we computed for each case the standard deviation ($\sigma$; 
indicative of the ``observation minus calculation residual) defined as
\begin{equation}
\sigma \equiv \sqrt{ 
\frac{ \sum_{i=1}^{N} [V_{\rm r}^{\rm obs}(t_{i}) + C - V_{\rm r}^{\rm cal}(t_{i})]^{2}}{N}.
}
\end{equation}
Here, $C$ is the offset value to be added to $V_{\rm r}^{\rm obs}$ to make the 
two means ($\langle V_{\rm r}^{\rm obs} \rangle$ and $\langle V_{\rm r}^{\rm cal} \rangle$) 
equal.\footnote{Note that we can apply any arbitrary constant to 
$V_{\rm r}^{\rm obs}$ since it is meaningful only in the relative sense.}
\begin{equation}
C \equiv 
\frac{ \sum_{i=1}^{N} [V_{\rm r}^{\rm cal}(t_{i}) - V_{\rm r}^{\rm obs}(t_{i})]}{N}.
\end{equation}

Since $\sigma (A, B)$ is a measure of goodness-of-fit between observation and calculation, 
what we should do next is to find the solutions of ($A$, $B$) that minimize $\sigma$. 
After inspection, however, we found it difficult to separately establish the best-fit 
values of these two parameters at the same time based on the behavior of $\sigma$ alone.
This situation is illustrated in figure 9, where we can see that the surface of
$\sigma (A, B)$ is like an elongated trough, in which pinpointing $(A, B)$ 
corresponding to $\sigma$-minimum is hardly possible. Actually, this is attributed to 
the fact that both $A$ and $B$ influence on the radial velocity curve in a similar manner.
That is, as manifested in figure 8, the effect of increasing $A$ (for a fixed $B$) 
is qualitatively the same as that of increasing $B$ (for a fixed $A$),
which means that an increase of $B$ can be compensated by an appropriate decrease
of $A$ without causing any appreciable change in the shape of $V_{\rm r}^{\rm cal}(t)$.
This is the reason for the difficulty of establishing $(A, B)$ from the radial-velocity alone.

Nevertheless, we can at least state that any combination of $(A, B)$ satisfying the equation
\begin{equation}
B \simeq -5.5 A + 77
\end{equation}
gives the overall best-fit between $V_{\rm r}^{\rm cal}(t)$ and $V_{\rm r}^{\rm obs}(t)$,
as shown in the contour diagram of figure 9.
Then, how is this relation compared with the real $(A, B)$ values known for the sun? 
According to seven representative studies (Ward 1966; Godoli \& Mazzucconi 1979;
Balthasar \& W\"{o}hl 1980; Ar\'{e}valo et al. 1982; Howard 1984; Balthasar et al. 1986;
Sivaraman et al. 1993) where the solar differential rotation was investigated based on 
the sunspot-tracing method, the resulting $(A, B)$ [in the definition of equation (3)] 
they determined were in good agreement with each other (cf. figure 12 of Takeda \& Ueno 2011), 
which give $\langle A \rangle = 14.56 (\pm 0.04)$ and  $\langle B \rangle = -2.81 (\pm 0.08)$
on the average (the value with $\pm$ in the parenthesis is the standard deviation).
Interestingly, these empirical values are remarkably consistent with equation (11),
since it yields $B = -2.75$ for $A = 14.5$. 
Actually, we can confirm in figure 10 that the $V_{\rm r}^{\rm cal}$ curve computed for 
$(A, B) = (14.5, -2.75)$ satisfactorily matches our $V_{\rm r}^{\rm obs}$ data.\footnote{
Precisely speaking, some systematic deviations are noticeable, especially 
in the first half of observation (before the mid-eclipse); e.g., the observed data are 
slightly below the theoretical curve at JD~2456068.40--2456068.41, while this tendency 
is inversed and the amplitude of $V_{\rm r}^{\rm obs}$ is somewhat insufficient at 
the velocity minimum of JD~2456068.43. Since the fit is more satisfactory for the 
second half of observation after the mid-eclipse, we suspect that the conditions before 
and after the mid-eclipse was not strictly symmetrical. 
It is interesting to note that a similar asymmetric tendency is observed in 
the comparison of theoretical and observed line-width described in the Appendix.
Accordingly, we interpret that this difference of $O-C$ in $V_{\rm r}$ 
before and after mid-eclipse (worse for the former and better 
for the latter) was caused because the crescent part being visible 
just before mid-eclipse happened to include some unusual region 
(presumably related to active phenomena) which could not be properly 
treated by our simple modeling.
In any event, since the solution is essentially determined by the latter half 
period after mid-eclipse where the fit is remarkably good, this slight 
$O-C$ inconsistency just before mid-eclipse does not cause any serious problem.
}
 
\subsection{Application of R--M effect to eclipsing binaries}

As we have shown above, we could successfully derive the realistic relation between
$A$ and $B$ (describing the nature of solar rotation) based on the radial velocity data 
of the sun-as-a-star (showing the typical R--M effect) alone, which were obtained 
during the 2012 May 21 solar eclipse, though separate determinations of $A$ and $B$ 
were not possible.  Then, how is the prospect of applying the R--M effect for 
studying the rotational properties of eclipsing binaries?

First, we would point out that useful information on the equatorial rotation velocity
($v_{\rm eq}$) can be obtained by using the R--M effect even if the nature of differential 
rotation is unknown, since the latter effect on the radial velocity curve is 
quantitatively less significant (e.g., compare figure 8b with figure 8a).
Let us suppose in the present case that $B$ is unknown and we had to assume 
the rigid rotation ($B =0$). Then, we would obtain $A = 14.0$ from equation (11),
which differs from the true $A$ value only by $\sim 4\%$.
Alternatively, it may be reasonable to assume $-B/A \simeq 0.2$ (solar case) for 
analyzing the R--M effect of general eclipsing binaries, which would not cause 
any serious error, since the differential rotation parameter of A--F stars studied by 
Ammler-von Eiff and Reiners (2012) (which they call $\alpha$, being equivalent 
to $-B/A$) ranges from $\sim 0.0$ to $\sim 0.5$.
Considering that spectroscopic determination of $v_{\rm eq} \sin i$ (though $i$ is 
often known and near $\sim 90^{\circ}$ for eclipsing binaries) with a precision 
of $\ltsim 10\%$ based on spectral line profiles is very difficult (especially 
for double-line binaries where spectral lines of both components are blended and 
spectra are appreciably time-variable), $v_{\rm eq}$ determination by utilizing 
the R--M effect would be valuable from the viewpoint of its accuracy.

On the other hand, we expect it much more difficult to get information on
the differential rotation based on the R--M effect alone. Of course,
determination of $B$ might be possible if $A$ (or equatorial rotation velocity) 
is precisely known in advance based on some other independent way, which 
however does not seem easy in general cases. Rather, we consider that this 
R--M effect may be useful to ``check'' the detailed rotational features 
established by other more sophisticated method such as the magnetic Doppler
imaging (making use of the spectrum variation caused by stellar rotation
and starspots) based on high-dispersion spectro-polarimetry (e.g., 
Donati \& Collier Cameron 1997; Collier Cameron \& Donati 2002).

Here, we should recall that many eclipsing binaries (especially close of binaries
of shorter periods) are tidally locked, so that the rotation and orbital motion  
are synchronized. If this is already known firmly, it may not be much meaningful 
to determine the rotational velocity or the differential parameter in an independent way. 
Yet, the accomplishment of co-rotation is not necessarily evident in general cases, 
especially for the case of young binaries, for which direct confirmation of 
the rotational parameters based on the R--M effect should still be meaningful. 
For example, in case where the tidal synchronization is realized, we should
obtain $A = 2\pi R/P_{\rm orbital}$ and $B = 0$. It would thus make the most 
definite evidence for the existence of co-rotation to confirm these two conditions.

Finally, it should be kept in mind that an R--M effect analysis such as done 
in this study presupposes an availability of observational data satisfying
rather high requirements (a series of many high dispersion spectra with 
sufficiently high S/N ratios,  which have to be obtained in sufficiently high
time resolution), since the time-variable radial velocity at each eclipse phase  
should be determined precisely.  Considering that many interesting eclipsing
binaries are comparatively faint (and that the integrated star light appreciably 
gets dimmer during the eclipse), these demands may appear somewhat too severe.    
Yet, such observations would become practically feasible when observing times 
of large 8--10~m class telescopes equipped with high-speed spectrometers 
are flexibly available.

\bigskip
We express our heartful thanks to Dr. Akihiko Fukui for his help
in preparing the observation.

E. K. is grateful for a financial support by the Grant-in-Aid for Scientific 
Research (No. 20540240) from the Japan Society for the Promotion of Science (JSPS).

N. N. acknowledges supports by NAOJ Fellowship, Inoue Science Research
Award, and Grant-in-Aid for Scientific Research (A) (No. 25247026) from 
the Ministry of Education, Culture, Sports, Science and Technology (MEXT).

\appendix
\section*{Behavior of spectral line width during eclipse}

As a supplementary analysis to our radial velocity study, we also examined 
based on our observational data how the width of a spectral line varies 
during the eclipse and whether it can be reproduced by theoretical simulations.
Since we want to measure the line width influenced by the Doppler motion
due to solar rotation, the spectral line should neither be too strong 
(to avoid large intrinsic width such as the case of a saturated line) 
nor too weak (to allow a sufficiently precise measurement).
Considering these requirements, we selected the Fe~{\sc i} line at 
6703.568~$\rm\AA$ ($\chi_{\rm low}$ = 2.76~eV)
for this purpose, which is in the wavelength region
almost unaffected by I$_{2}$ lines (cf. figure 4b).

We computed the intrinsic line profile (of specific intensity) at each point 
of the solar disk (where the same division was used as in figure 7)
by using Kurucz's (1993) ATLAS9 solar model [$T_{\rm eff}$ = 5780~K (effective temperature ), 
$\log g = 4.44$ (logarithmic surface gravity in cm~s$^{-2}$),
$v_{\rm t}$ = 1~km~s$^{-1}$ (microturbulence), [Fe/H] = 0 (metallicity)], 
which was further convolved with the radial-tangential mactoturbulence
function (parameterized with $\zeta_{\rm R}$ and $\zeta_{\rm T}$; cf. Gray 2005)
as well as the Gaussian instrumental profile corresponding to $R = 52000$,
and finally integrated over the visible disk to get the final profile.  
In this computation, we adopted an adjusted (astrophysical) $\log gf$ value of 
$-2.90$; that is, Kurucz and Bell's (1995) original $\log gf$ of $-3.16$ was 
increased by +0.26~dex, so that the observed line strength can be reproduced 
for the solar Fe abundance of $\log \epsilon = 7.50$.
Regarding the macroturbulence parameters, we assumed 
$\zeta_{\rm R} = 2$~km~s$^{-1}$ and $\zeta_{\rm T} = 3$~km~s$^{-1}$
by consulting figure 2 of Takeda (1995), while considering that
the mean-forming depth of this line is $\log \tau_{5000} \sim -1$.
As to the combination of $(A, B)$, three representative cases satisfying 
equation (11) were simulated: (13.0, +5.5), (14.0, 0.0), and (15.0, $-5.5$).

We then calculated the FWHMs for each of the resulting theoretical profiles  
corresponding to various eclipse phases, which are to be compared with
the observed FWHMs directly measured from our 184 spectra.
Such comparisons are displayed in figure 11a, where we can see that
the line width shows a variation with the following characteristics: \\
--- At the ingress phase, it first decreases gradually until attaining 
a minimum width around $\sim$~JD~245668.42 where only the approaching 
east limb is visible (near to the minimum $V_{\rm r}$).\\
--- Then it turns to grow as the receding west limb begins 
to reappear, and eventually shows a maximum sharp peak at the mid-eclipse
(at JD~245668.44) where both of the approaching east limb and the receding 
west limb are equally visible.\\
--- After that, in the egress phase, the width decreases abruptly and again becomes 
minimum around $\sim$~JD~245668.46 when only the receding west limb 
is visible (near to the maximum $V_{\rm r}$), then begins to gradually
increase toward the eclipse end.

It is immediately noticed from figure 11a that the mutual differences 
between the FWHM curves computed for three $(A, B)$ combinations are
very subtle and hardly discernible. Thus the situation is quite the same
as the case of the radial velocity, meaning that separate determinations 
of $(A, B)$ based on the FWHM analysis are not possible.

According to the comparison shown in this figure, our observed FWHM data surely 
exhibit such an expected qualitative tendency, though not necessarily satisfactory
in the quantitative sense. Especially, the deviation is appreciable at the 
phases before the mid eclipse (JD~2456068.4--2456068.3), while a more or less
tolerable consistency is observed after the mid-eclipse. It is interesting
that this asymmetric situation is somewhat similar to the case of figure 10
as mentioned in subsection 4.2 (footnote 12).
We also point out that the line depth is expected to become slightly larger
at the mid-eclipse compared to the value outside of the eclipse according to 
our calculation, which does not observed in our spectra (cf. figure 11b). 
This may imply that our line-profile modeling needs to be further improved
if much better fitting is to be pursued.

\newpage
\onecolumn

\begin{figure}
  \begin{center}
    \FigureFile(140mm,140mm){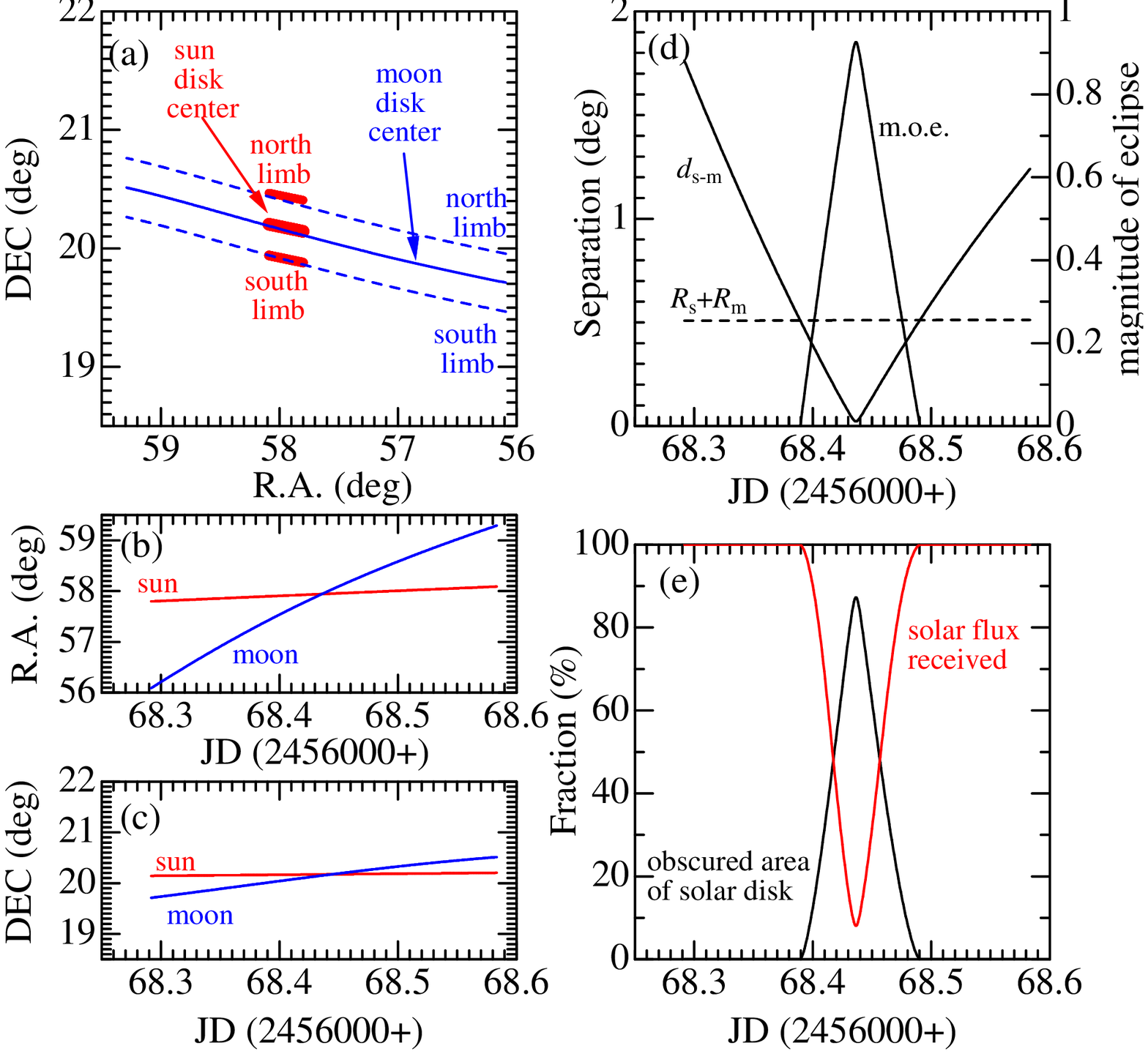}
  \end{center}
\caption{
Graphical description of the solar eclipse at Okayama on 2012 May 21.
(a) Tracks of the sun and moon on the celestial sphere. (b) Change of right ascension 
with time. (c) Change of declination with time. (d) Change of sun--moon separation 
($d_{\rm s-m}$) and the magnitude of eclipse (m.o.e.), which is defined as 
$(R_{\rm s} + R_{\rm m} - d_{\rm s-m}) / (2 R_{\rm s})$, with time. 
(e) Fraction of the sun's obscured area and that of the total received flux 
(evaluated by the simple limb-darkening model of equation (8)), plotted against time.
}
\end{figure}

\begin{figure}
  \begin{center}
    \FigureFile(120mm,150mm){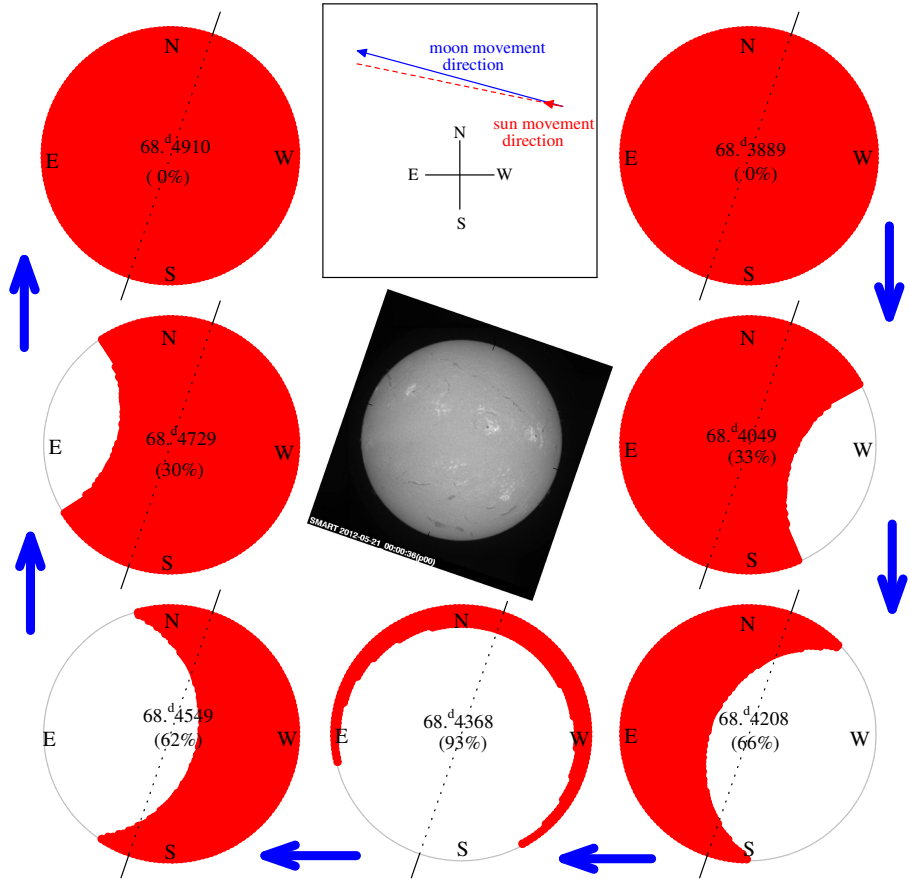}
  \end{center}
\caption{
Schematic display of how the sun was obscured at seven representative phases
of the 2012 May 21 solar eclipse observed at Okayama, where the 
corresponding Julian day (+2456000) as well as the degree of eclipse are 
indicated. The direction of the solar rotation axis is also marked on each disk.
For a reference, the H$\alpha$(-center) filtergram on this day at 00:00:16 UT
(just after the ending of eclipse) observed by SMART in Hida observatory
of Kyoto University is also presented, in order to indicate the nature
of active regions on the solar surface. 
Shown in the inset are the directions for the movement of the sun and moon 
on the celestial sphere (the latter is by 10.9 times faster than the former). 
}
\end{figure}

\begin{figure}
  \begin{center}
    \FigureFile(120mm,180mm){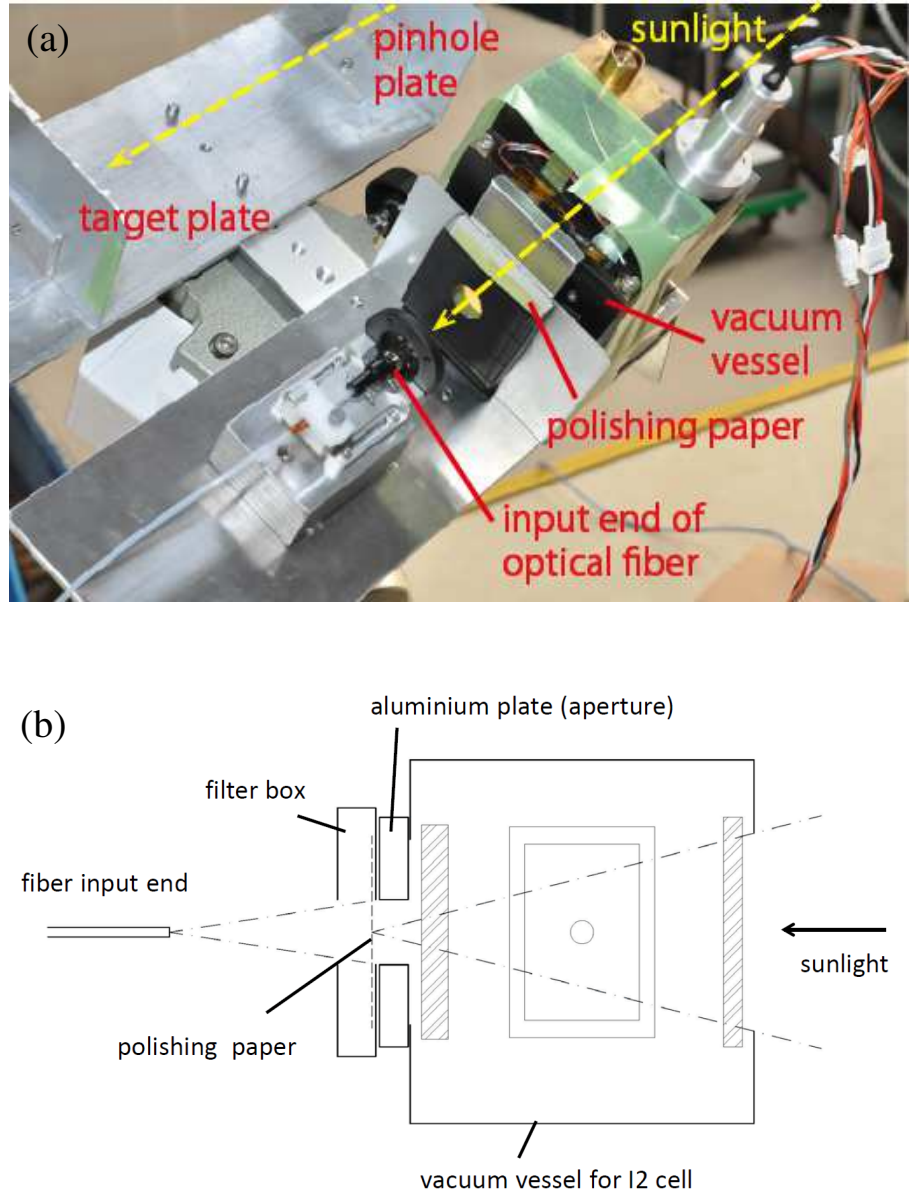}
  \end{center}
\caption{
(a) Photographic view of the fiber entrance components on the portable telescope mount.
(b) Schematic optical layout showing how the incident sunlight passes through 
the I$_{2}$ cell and reach the fiber (leading to the spectrograph).
See subsection 2.1 for a more detailed explanation.
}
\end{figure}

\begin{figure}
  \begin{center}
    \FigureFile(100mm,150mm){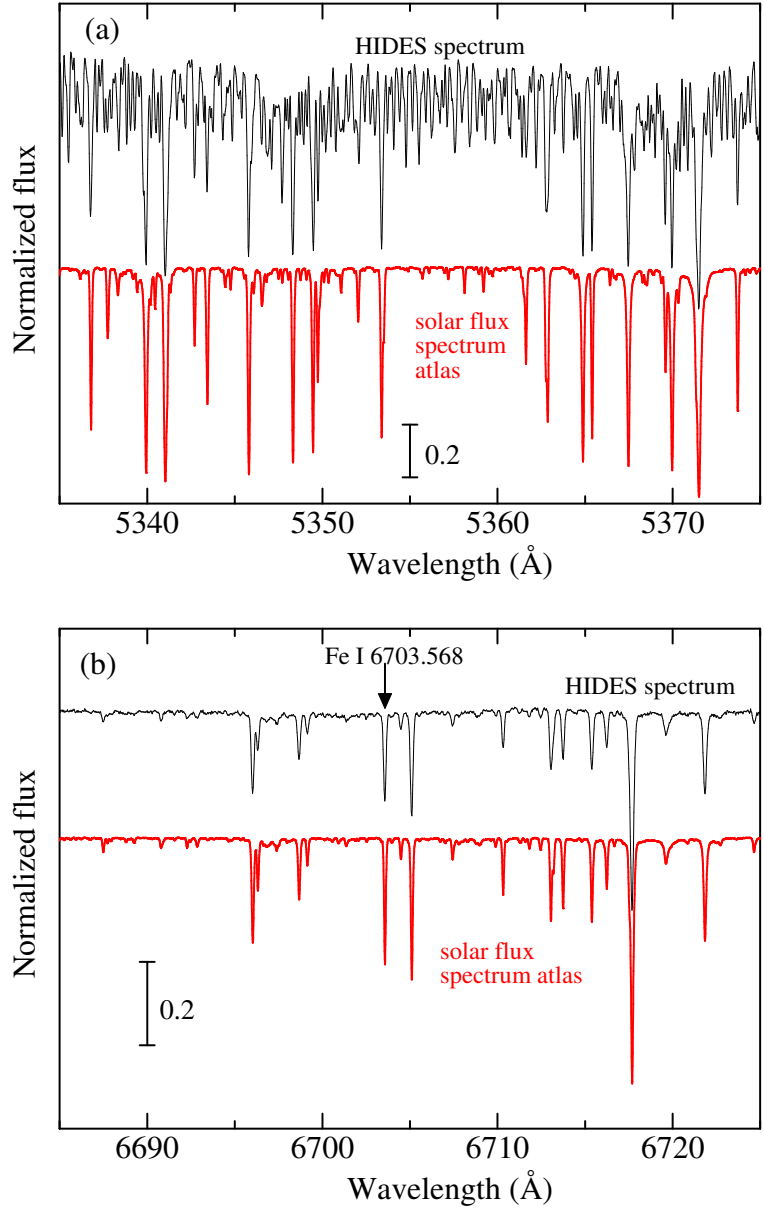}
  \end{center}
\caption{
Typical spectra which we observed with HIDES at JD 2456068.49 (just outside 
of the eclipse). (a) 5335--5375~$\rm\AA$ region (where absorption features 
of many I$_{2}$ molecular lines caused by the iodine cell are manifest). 
(b) 6685--6725~$\rm\AA$ region (where I$_{2}$ lines are almost absent).  
For reference, the solar flux spectra taken from Kurucz et al.'s (1984) atlas are also shown.
In panel (b), the position of the Fe~{\sc i} 6703.568 line, for which the 
line profile was simulated (cf. the Appendix), is also indicated.
}
\end{figure}

\begin{figure}
  \begin{center}
    \FigureFile(120mm,120mm){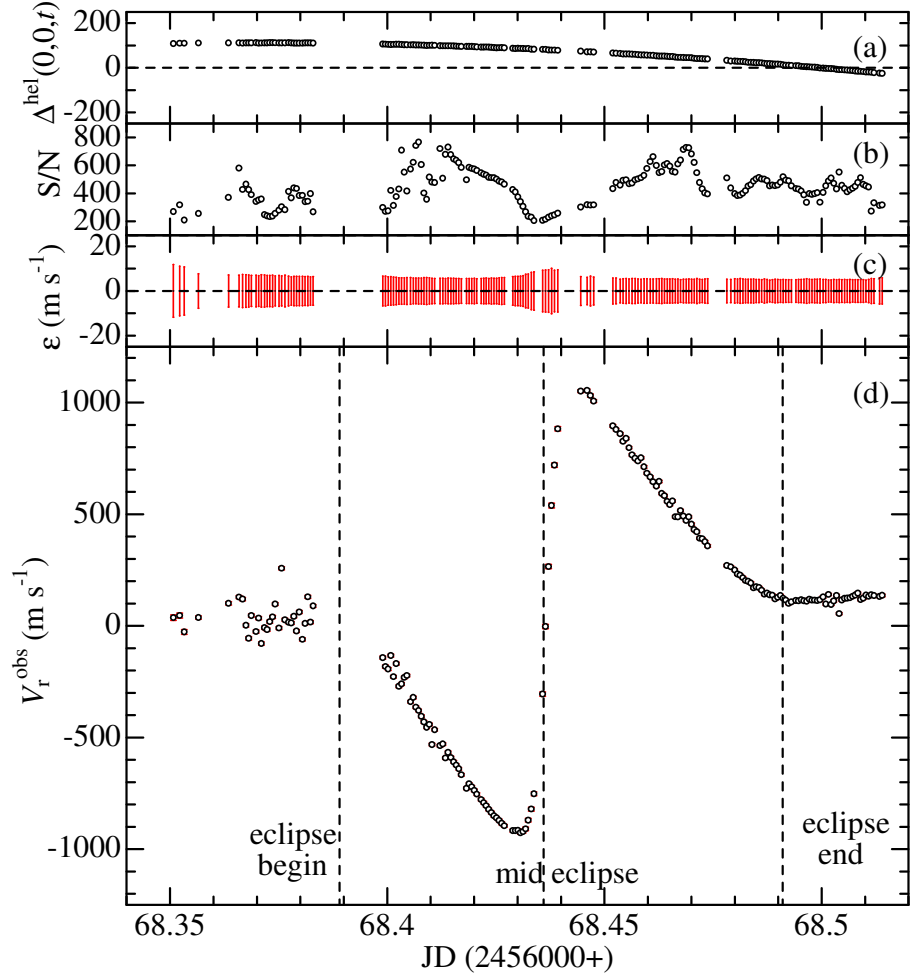}
  \end{center}
\caption{
Run of the observation-related quantities with time.
(a) Heliocentric correction due to earth's motion ``at the sun's disk center'' 
($\Delta^{\rm hel}(0,0,t)$; note that this can not be simply applied to
$V_{\rm r}^{\rm obs}$, as remarked in footnote 9).
(b) S/N ratio of each spectrum estimated from the photon count at $\sim 5500$~$\rm\AA$. 
(c) Estimated error ($\pm \epsilon$) involved with each radial velocity
[note that the ordinate scale is 10 times expanded compared with panels (a) and (d)].
(d) Observed (uncorrected) radial velocity ($V_{\rm r}^{\rm obs}$ 
relative to the template (including an offset constant of $-845.5$~m~$s^{-1}$),
where the error bars [depicted in panel (c)] are attached to the symbols
but hardly discernible. Note that the velocity scale in the ordinate of 
panel (c) is by 10 times expanded as compared to panels (a) and (d). 
}
\end{figure}

\begin{figure}
  \begin{center}
    \FigureFile(80mm,80mm){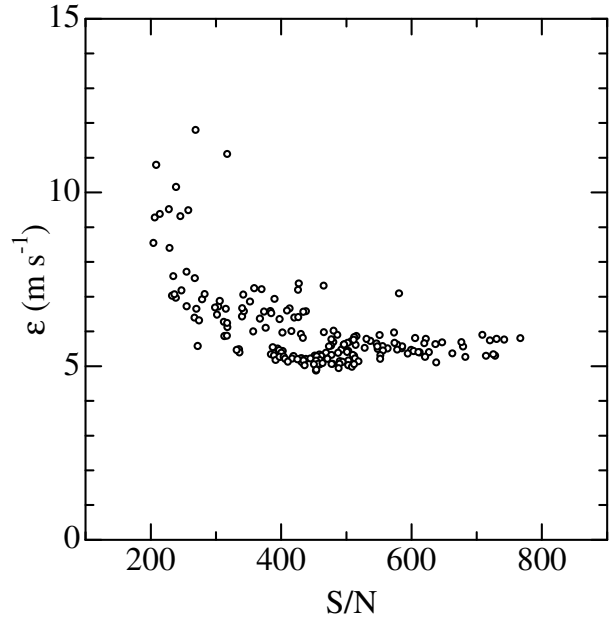}
  \end{center}
\caption{
Correlation between the S/N ratio of each spectrum and the error
($\epsilon$) involved with the derived radial velocity.  
}
\end{figure}

\begin{figure}
  \begin{center}
    \FigureFile(80mm,160mm){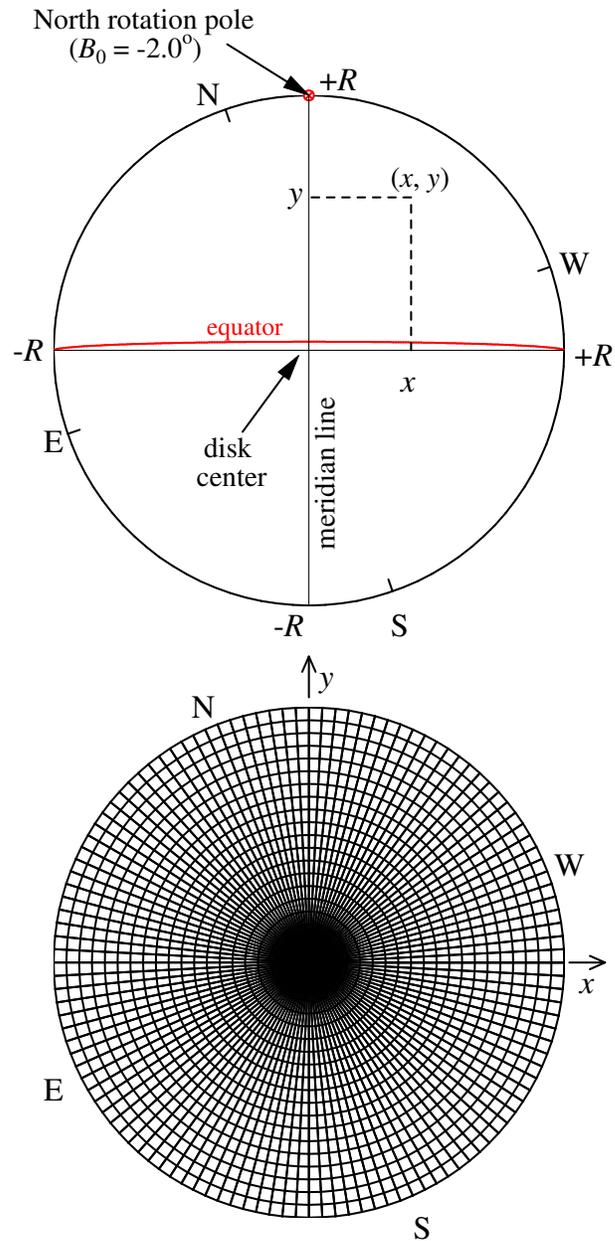}
  \end{center}
\caption{
Coordinate setting and segmentization of the solar disk for simulating
the radial velocity (and spectral line profile). Note that the actually adopted 
division ($100 \times 360$ with steps of $0.01 R$ and 1~deg) is by 
$5 \times 3$ times finer than that shown here ($20 \times 120$ with steps of 
$0.05 R$ and 3~deg; just to avoid too much complexity). 
}
\end{figure}

\begin{figure}
  \begin{center}
    \FigureFile(120mm,150mm){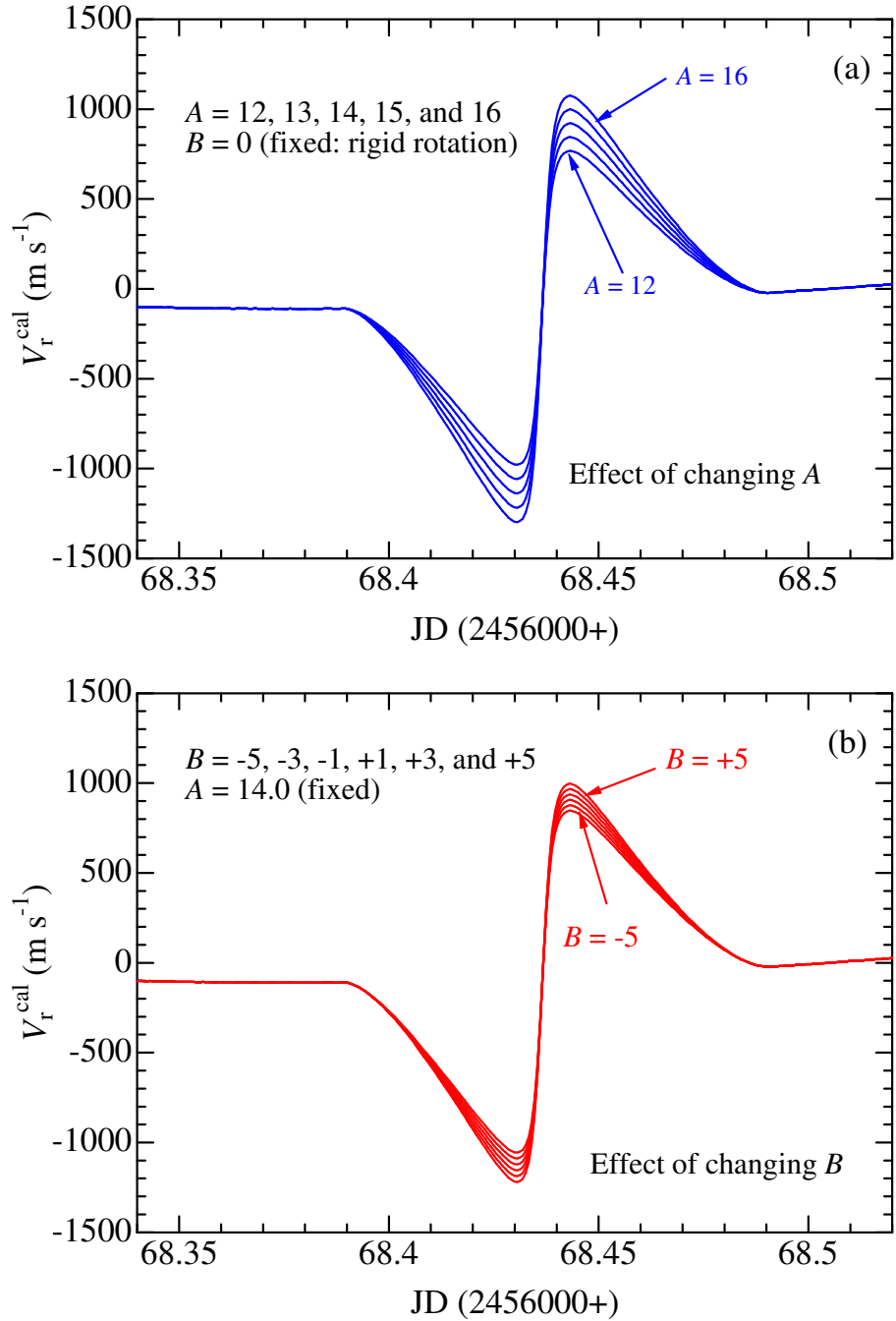}
  \end{center}
\caption{
Run of the theoretically modeled radial velocity with time during the eclipse.
(a) Results for five different $A$ values (12, 13, 14, 15, and 16) 
while $B$ is fixed at 0.0 (rigid rotation). (b) Results for six different $B$ 
values ($-5$, $-3$, $-1$, +1, +3, and +5) while $A$ is fixed at 14.0.
}
\end{figure}

\begin{figure}
  \begin{center}
    \FigureFile(120mm,150mm){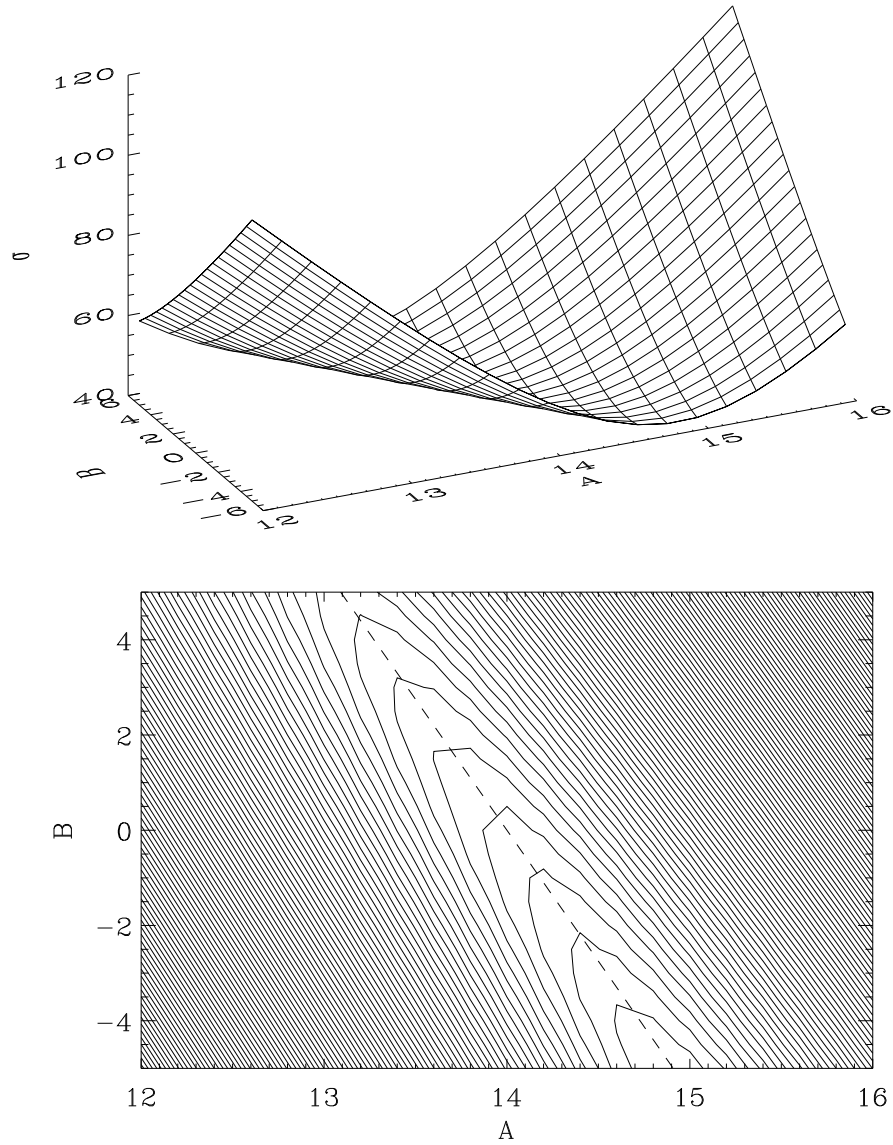}
  \end{center}
\caption{
Graphical representation of the behavior of $\sigma(A,B)$ (standard deviation 
in fitting the simulated radial velocity curve with the observed data).
The upper panel is the 3D representation of the $\sigma(A,B)$ surface,
while the lower panel shows the contours of $\sigma$ on the $A$--$B$ plane,
where the relation $B = -5.5 A + 77$ is drawn by dashed line.  
The unit of $\sigma$ is m~s$^{-1}$, while that of $A$ and $B$ is deg~day$^{-1}$.
}
\end{figure}

\begin{figure}
  \begin{center}
    \FigureFile(120mm,120mm){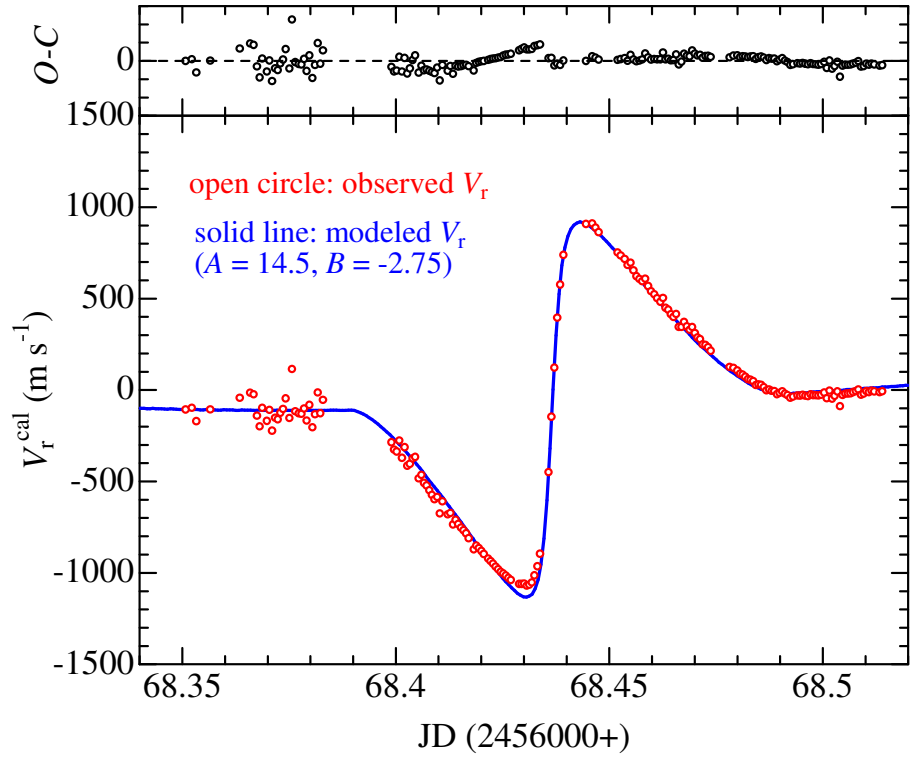}
  \end{center}
\caption{
Demonstration that how the combination of ($A =14.5$, $B = -2.75$), which is
not only consistent with our derived relation (equation (11)) but also 
the real solution for the actual solar rotation, accomplishes the
satisfactory match between the calculation (solid line) and observation (open symbols).
In this case, the offset for $V_{\rm r}^{\rm obs}$ is $C = -143$~m~s$^{-1}$
and the standard deviation is $\sigma = 41$~m~s$^{-1}$.  
}
\end{figure}

\begin{figure}
  \begin{center}
    \FigureFile(120mm,140mm){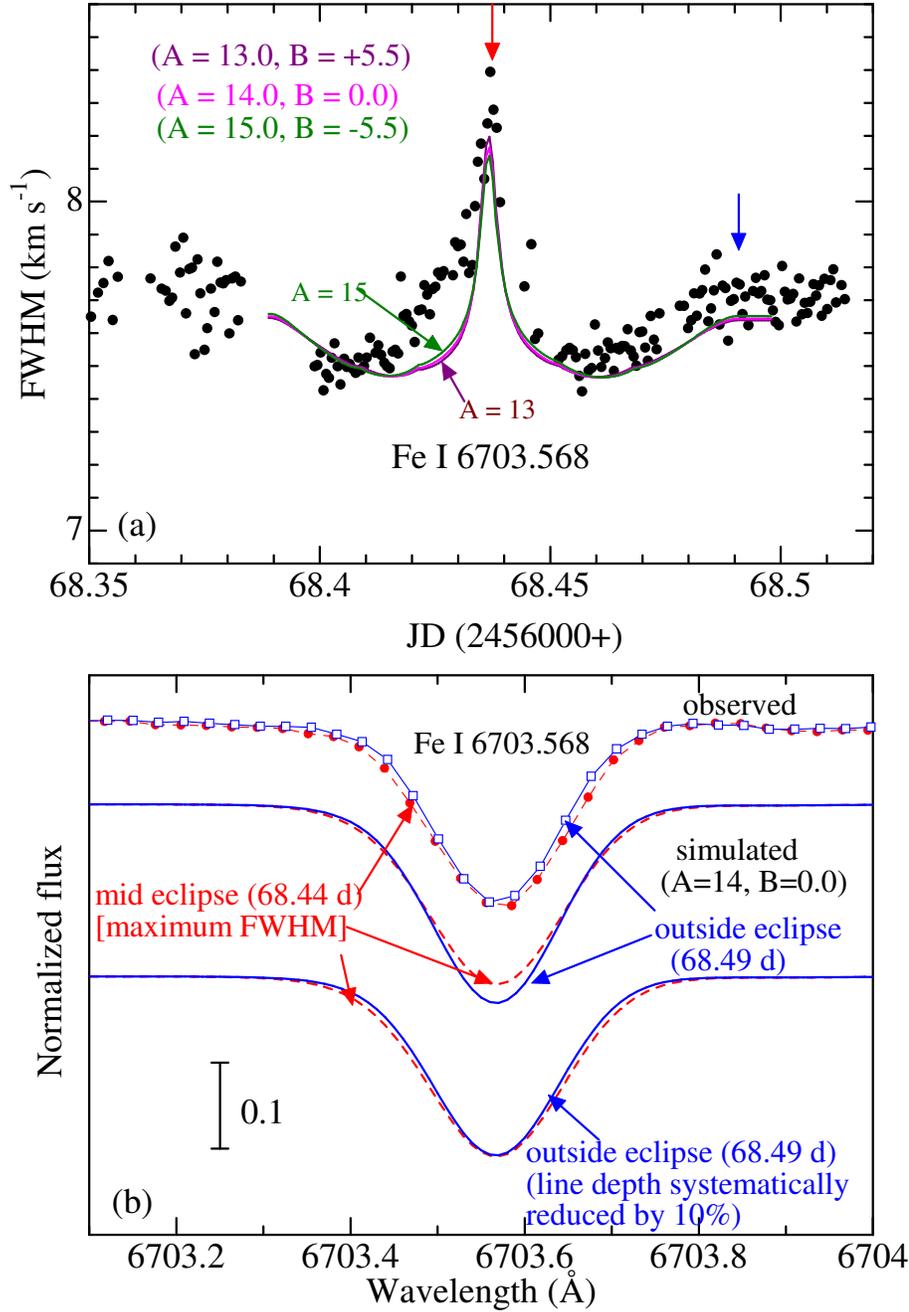}
  \end{center}
\caption{
(a) Comparison of the observed (symbols) and simulated (lines) FWHM of Fe~{\sc i} 6703.568 
during the course of eclipse. The modeled curves are computed with three different
($A$, $B$) combinations of (13.0, +5.5), (14.0, 0.0), and (15.0, $-5.5$)
satisfying equation (11). (b) Comparison of the  Fe~{\sc i} 6703.568 line profiles
at two representative phases indicated by downward arrows in panel (a) (mid-eclipse 
with the largest FWHM at JD~2456068.44 and outside of eclipse at JD~2456068.49). 
Top: observed profiles (line-connected filled and open symbols correspond to 
mid- and outside-eclipse, respectively). 
Middle: theoretical profiles (the dashed and solid line corresponds to mid- and 
outside-eclipse, respectively).
Bottom: almost the same as the middle, but the line depth of the stronger one
(solid line) is intentionally reduced by 10\% in order to make the direct 
comparison of both FWHM easier.
}
\end{figure}

\end{document}